# Quantitative Comparisons of Linked Color Imaging and White-Light Colonoscopy for Colorectal Polyp Analysis

Xinran Wei[1], Jiyang Xie[1], Wenrui He[1], Min Min[2,*], Zhanyu Ma[1,*], and Jun Guo[1]

[1] Pattern Recognition and Intelligent Systems Lab, Beijing University of Posts and Telecommunications, Beijing, China

[2] Department of Gastroenterology and Hepatology, Affiliated Hospital of Academy of Military Medical Sciences, Beijing, China

**Abstract:** The performance of imaging techniques has an important influence on the clinical diagnostic strategy of colorectal cancer. Linked color imaging (LCI) by laser endoscopy is a recently developed techniques, and its advantage in improving the analysis accuracy of colorectal polyps over white-light (WL) endoscopy has been demonstrated in previous clinical studies. However, there are no objective criteria to evaluate and compare the aforementioned endoscopy methods. This paper presents a new criterion, namely entropy of color gradients image (ECGI), which is based on color gradients distribution and provides a comprehensive and objective evaluating indicator of the performance of colorectal images. Our method extracts the color gradient image pairs of 143 colonoscopy polyps in the LCI-PairedColon database, which are generated with WL and LCI conditions, respectively. Then, we apply the morphological method to fix the deviation of light-reflecting regions, and the ECGI scores of sample pairs are calculated. Experimental results show that the average ECGI scores of LCI images (5.7071) were significantly higher than that of WL (4.6093). This observation is consistent with the clinical studies. Therefore, the effectiveness of the proposed criterion is demonstrated.

**Keywords:** Linked color imaging; colorectal polyp; color gradient; quantifiable evaluating indicator

## 1 INTRODUCTION

Colorectal cancer (CRC) is the third most commonly diagnosed cancer among both men and women in the United States [1]. The colonic polyps have a rather high prevalence and are known to be precursors to colon cancer [2]. Colonoscopy is an effective method for detecting and removing adenomatous polyps and has resulted in significant decrease in the incidence and the mortality of colon cancer [3]. However, the quality of colonoscopy depends on human and instrumental factors and it affects the polyp miss and misclassification rate considerably [4]. Patients with missed adenomatous polyps may later be diagnosed with advanced colorectal cancer with a survival rate of less than 10% [5].

Great progress has been made on the endoscopic technique in order to improve the diagnostic accuracy and therapeutic efficacy [6] [7]. Chromoendoscopy and magnified endoscopy, among others, have the most significant contribution, while they are time consuming and are not widely used in most parts of the world [8]. A few techniques of endoscopy such as white-light (WL) and linked color imaging (LCI) have been applied in medical equipment. LCI by laser endoscopy is a novel narrow band light observation, which uses a laser endoscopic system to enhance the color separation of red color to depict red and white colors more vividly [9]. Research on different endoscopy will optimize the current clinical strategy and will provide more professional assistance for doctors to differentiate between the mosaic and the normal mucosa [10].

There are two clinical tasks for colorectal polyp analysis: detection and classification. In recent research in the field of medical science, a subjective marking criterion and a diagnostic accuracy criterion were applied to evaluate colonoscopy performance in both tasks. In the subjective marking criterion, the experts scored the visibility and color difference of the colorectal image according to their subjective feelings [11]. As for another method, two experiments were designed according to the diagnoses of endoscopists combined with histological verification. Two sets of results based on pathology including random crossed experiments [12] for detection and diagnostic accuracy experiments for classification [13] are showed in Table I. In both experiments, the miss and misclassification rates under LCI colonoscopy are an order of magnitude lower than that of WL, respectively, which demonstrated the clinical benefits of LCI colonoscopy in clinical adjuvant diagnosis, compared with WL endoscopy.

**Table I** Results of clinical methods [12, 13]

| Task | Criterion | LCI | WL | *P-value* |
|---|---|---|---|---|
| Detection [12] | miss rate | 3.41% | 22.90% | <0.001 |
| Classification [13] | misclassification rate | 4.50% | 54.50% | <0.001 |

However, there are still lacks of comprehensive quantifiable criteria for polyp analysis. The previous subjective comparisons reflect a particular characteristic, and it is uncertain depending on the subject group. Although the method based on pathological diagnosis is the best standard [14], it costs a lot of time and labor work while it is unreviewable. Therefore, existing methods are not suitable for the generalized research on colorectal images in the testing phase of medical devices. Computer-aided methods [15] developed rapidly under the basic theory such as variational Bayes [16] [17] [18], and they are increasingly used in the medical field. In this paper, we propose a quantitative method based on the color gradients for comparing the performance of LCI and WL colonoscopy in analysis of polyps.

The rest of the paper is organized as follows: In Section 2, we describe the implementation of entropy of color gradients image (ECGI) criterion in detail. In Section 3, we give an introduction to our LCI-PairedColon database. The experimental results on the database and related discussions are presented in Section 4. Finally, we draw our conclusions in Section 5.

## 2 METHODOLOGY

### 2.1 Problem Formulation

In the medical domain, better performance generally means that lower polyps miss and misclassification rate. In this paper, we describe the problem of comparisons as the following statements. For every single input image $X$, we extract the feature $G$ and mark it with a score $S$. In sample pairs, the one with the higher score is considered with better performance.

For efficiency diagnosis, the texture, the edge, and the color characteristics are the major concerns for experienced endoscopists and polyp detection algorithms [19]. To simplify this process, we need to introduce a feature with the advantage that the color information of colonoscopy can be better preserved while the texture definition and the edge sharpness are under consideration as well. Therefore, we suppose the color gradient image is an essential feature and its score has a deterministic relation with performance.

### 2.2 Implement Solution

In this study, method proposed in [20] is applied to the color gradient calculation. For any pixel located in $(x, y)$ in the image as Figure 1(a), we define vector gradients $u$ and $v$, and their dot product $g$ as

$$u = \left[\frac{\partial R}{\partial x}, \frac{\partial G}{\partial x}, \frac{\partial B}{\partial x}\right]^{\mathrm{T}}, \quad (1)$$

$$v = \left[\frac{\partial R}{\partial y}, \frac{\partial G}{\partial y}, \frac{\partial B}{\partial y}\right]^{\mathrm{T}}, \quad (2)$$

$$g_{xx} = u^{\mathrm{T}} \cdot u = \left|\frac{\partial R}{\partial x}\right|^2 + \left|\frac{\partial G}{\partial x}\right|^2 + \left|\frac{\partial B}{\partial x}\right|^2, \quad (3)$$

$$g_{yy} = v^{\mathrm{T}} \cdot v = \left|\frac{\partial R}{\partial y}\right|^2 + \left|\frac{\partial G}{\partial y}\right|^2 + \left|\frac{\partial B}{\partial y}\right|^2, \quad (4)$$

$$g_{xy} = u^{\mathrm{T}} \cdot v = \frac{\partial R}{\partial x}\frac{\partial R}{\partial y} + \frac{\partial G}{\partial x}\frac{\partial G}{\partial y} + \frac{\partial B}{\partial x}\frac{\partial B}{\partial y}, \quad (5)$$

where $R$, $G$, and $B$ are image components of red, green, and blue color channel, respectively. Thus, gradient direction $\theta(x, y)$ of location $(x, y)$ is as

$$\theta(x, y) = \underset{\theta}{argmax}[|u\cos\theta + v\sin\theta|^2]$$

$$= \frac{1}{2}\tan^{-1}\left[\frac{2g_{xy}}{g_{xx}-g_{yy}}\right]. \quad (6)$$

Then, the corresponding gradient is as

$$F(x,y) = \left\{\frac{1}{2}\left[(g_{xx} + g_{yy}) + (g_{xx} - g_{yy})\cos 2\theta(x,y) + 2g_{xy}\sin 2\theta(x,y)\right]\right\}^{\frac{1}{2}}. \quad (7)$$

The color gradient amplitude image $F$ as Figure 1(b) can be drawn by calculating the value of each pixel.
Partial overexposure caused by particular lighting condition is inevitable during enteroscopy and these glistening spots can lead to false high gradients. Removing their pixels directly from polyp images will exert a terrible effect on adjacent gradients. Thus, in this paper, we propose a method to filtrate light-reflecting regions based on morphological method on the color gradient image, and replace them with a complemental value of gradient. Then, we apply the classical Maximally Stable Extremal Regions (MSER) algorithm [21] to detect the connected regions of the color gradient image. Only pixels with a gradient value greater than 0.2 are considered valid in order to avoid the error connection, and the area size of the connected regions is limited in the range of [5, 200]. Then, we plot all connected areas detected as a grayscale image as Figure 1(c), perform morphological closure [20] on the images and do connected component analysis. Through the properties of the connected regions, we screen out light-reflecting regions.

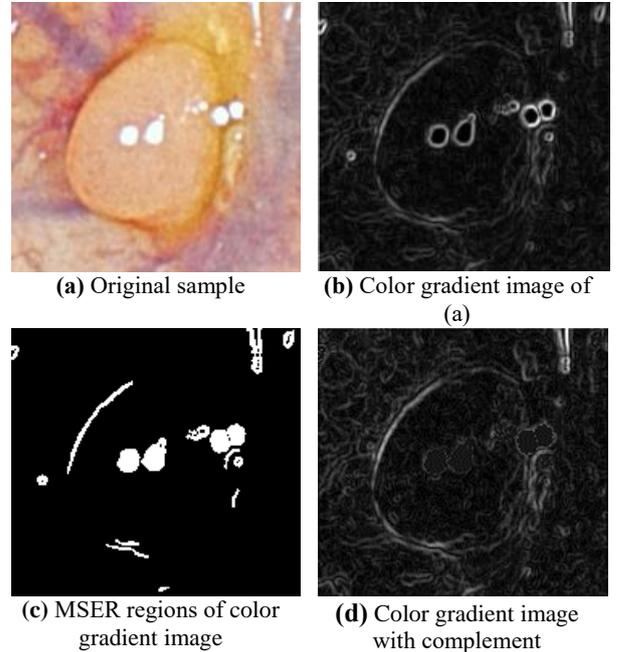

(a) Original sample (b) Color gradient image of (a)

(c) MSER regions of color gradient image (d) Color gradient image with complement

**Figure 1** A series of colorectal images during ECGI score calculation.

Make the values of color gradient in light-reflecting regions equal to complemental value, then we have the final color gradient image $G$ as Figure 1(d), and each pixel is computed by

$$G(x,y) = \begin{cases} c & (x,y) \in C \\ F(x,y) & otherwise \end{cases} \quad (8)$$

where $C$ is the connected regions of $F$, the complemental value $c$ of a color gradient image was described as

$$c = \mathbb{E}[F(x,y)], \quad (9)$$
$$s.t. \quad (x,y) \in F, \ 0 < F(x,y) < 0.2$$

where $\mathbb{E}[\cdot]$ is the expectation operator. The score $S$ of $G$ is defined as its entropy, that is

$$S = -\sum_n P_n \log_2 P_n, \quad (10)$$

where $P_n$ is $n^{th}$ bin of the probability mass function (PMF, $\boldsymbol{P} = [P_1, \cdots, P_{256}]^{\mathrm{T}}$) of $G$ which is quantized into 256 equal parts.

For a test set with $N$ pairs of samples ($N$ is a constant),

we calculate $S^{(LCI)}$ and $S^{(WL)}$ for both series of color gradient images under LCI and WL colonoscopy. $S^{(LCI)}$ and $S^{(WL)}$ are vectors as

$$S^{(LCI)} = \left[S_1^{(LCI)},\cdots,S_N^{(LCI)}\right]^T, \quad (11)$$

$$S^{(WL)} = \left[S_1^{(WL)},\cdots,S_N^{(WL)}\right]^T. \quad (12)$$

## 3 DATABASE

A multicenter, crossover, coupled, randomized collection was performed in several hospitals in China especially in the 307 Hospital of Academy of Military Medical Science for LCI-PairedColon database. Hundreds of patients underwent crossover colonoscopies with LCI and WL endoscopy in a randomized order.

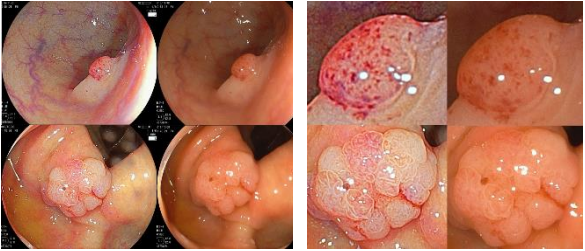

**(a)** Two original colorectal image pairs    **(b)** Corresponding extracted ROI pairs of (a)

**Figure 2** Comparison of colorectal image pairs before and after pre-processing (Notation: left: LCI, right: WL).

The LCI-PairedColon database consists of 143 image pairs (286 images) screening 143 polyps (hyperplastic and adenomatous polyps). An images pairs were taken orbiting around each polyp under ethical approval, where one is under WL endoscopy and the other one is under LCI. Two collected LCI images are shown on the left side in Figure 2(a), and the WL images in corresponding position on their right side. For each image, we invited experienced endoscopists to classify the polyp positions which were confirmed by histology, and considered as the ground truth.

All images were taken under random lighting conditions and zooming. Magnification endoscopy is not used in this study. Moreover, the size of the polyps and the background of the images have large variations. The collected endoscopic images have the same aspect ratio but three different resolutions: 1280 × 1024, 900 × 720, and 720 × 576. LCI-PairedColon database is available at https://github.com/weixinran/LCI-PairedColon-database.

For fair comparison, pre-processing is necessary for removing unwanted difference such as background. In this paper, 143 ROI (Region of Interests) pairs of each image are manually extracted from samples of this dataset according to the marked ground truths. Figure 2(b) shows two extracted ROI pairs of the images from the corresponding position in Figure 2(a). All ROIs are set to rectangles which center on polyps and contain polyp edge and have the same size.

## 4 EXPERIMENTAL RESULTS AND DISCUSSIONS

### 4.1 Experimental Results

A total of 143 ROI pairs in LCI-PairedColon database are analyzed in this study. According to the method proposed above, $S^{(LCI)}$ and $S^{(WL)}$ of the 143 pairs of samples are calculated, respectively for LCI and WL.

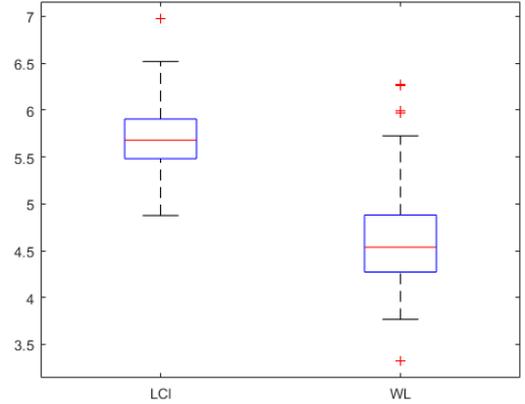

**Figure 3** Comparison of the distributions of ECGI scores. The central mark is the median, the edges of the box are the 25th and 75th percentiles. The outliers are marked individually.

The statistical distribution and of $S^{(LCI)}$ and $S^{(WL)}$ is shown in the boxplot of Figure 3. The ladder diagram of Figure 4 shows ECGI scores of images pairs, respectively, where ECGI scores of LCI are in bold. The ECGI scores of LCI images are significantly different from WL on a numerical scale. On the whole, the trapezoidal line of LCI is always higher than WL.

**Table II** Comparisons of the ECGI scores

| LCI (mean) | WL (mean) | P-value | Percentage (LCI>WL) |
|---|---|---|---|
| 5.7071 | 4.6093 | <0.0001 | 96.5% |

Table II shows the statistical results of LCI and WL on the ECGI scores of all the 143 ROI pairs. These results indicate that the average score of LCI color gradient images is larger than that of WL as Table II. Therefore, we find that the color gradient images of LCI have larger average information entropy. Also, the *p-value* obtained from paired *t*-test proves that there are significant differences between the scores of LCI and WL. Meanwhile, 138 (96.5%) scores of $S^{(LCI)}$ are larger than the corresponding $S^{(WL)}$, which demonstrates that the superiority of the LCI ECGI scores is robust.

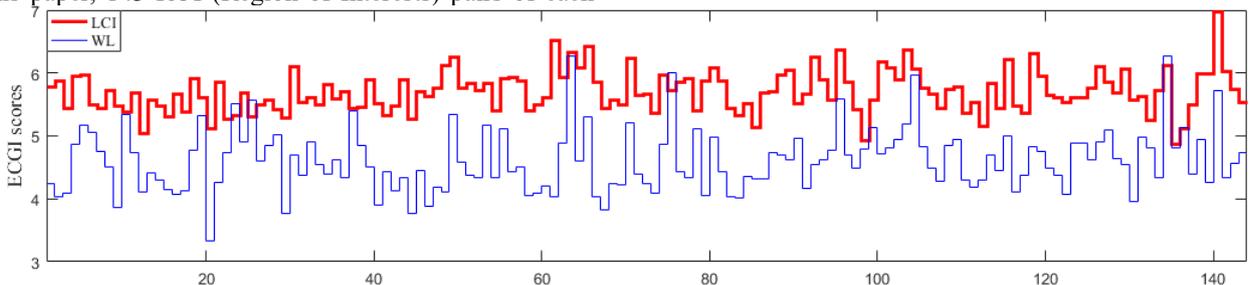

**Figure 4** Comparison of ECGI scores of LCI and WL images.

### 4.2 Discussions

According to the aforementioned results, ECGI is highly correlated with the tumor information. It makes the difference between LCI and WL striking. In terms of the proposed ECGI criterion, our results show the superiority of the samples under LCI colonoscopy than WL.

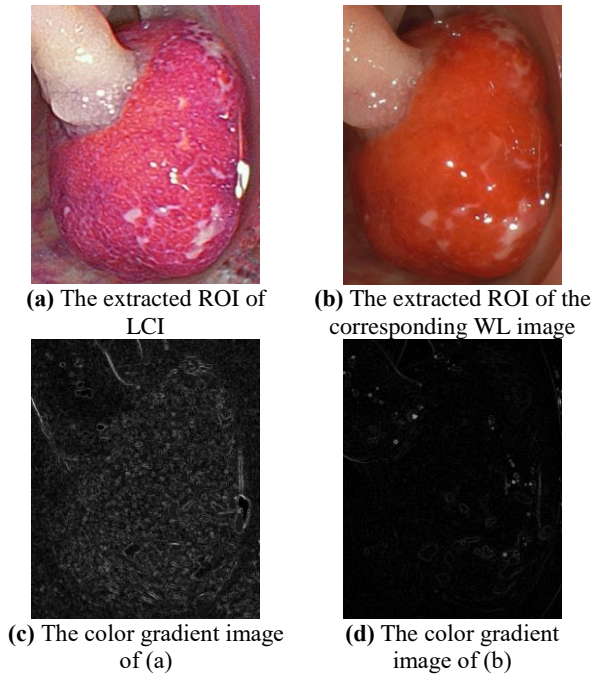

(a) The extracted ROI of LCI
(b) The extracted ROI of the corresponding WL image
(c) The color gradient image of (a)
(d) The color gradient image of (b)

**Figure 5** Comparison of color gradient image pairs.

Figure 5 shows a pair of colorectal images and their color gradient images as an example to explain the effectiveness of our ECGI criterion. It is obviously to see that the colorectal image in Figure 5(a) has more detailed information by its more obvious edges which are shown in Figure 5(c). The corresponding probability histograms of the color gradient images in Figure 5(c) and 5(d) are shown in Figure 6(a) and 6(b), respectively. Note that a probability histogram of a color gradient image is usually a unimodal distribution. Moreover, the gradient values of all points are distributed in the interval of 0 to 0.33. In this case, larger ECGI means the larger entropy of the color gradient image. Thus, the distribution of LCI color gradients is more flat (close to a uniform distribution), which indicates that there are more pixels with larger gradients in the LCI image than WL. Moreover, it also indicates that the LCI colorectal image has more texture, edge, and color information of the origin colorectal image. This conclusion is consistent with observation results by experts.

In clinical practice, the polyps miss and misclassification rate is considered the sole criterion for the comparison of colonoscopy [22]. In order to assess the accuracy of the proposed ECGI criterion in the colonoscopy, we summarize the results of medical comparisons in Table I and observe their consistency with our results. Although the experiments are performed on different datasets, our results show a high degree of consistency with the medical ones.

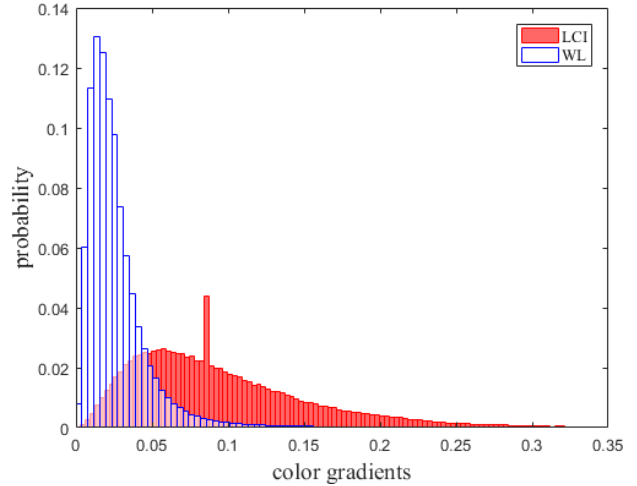

**Figure 6** Comparison of the distributions of color gradients in Figure 5(c) and 5(d)

Given the above observation, the color gradient image has been demonstrated to be a suitable feature and its entropy is considered to have a deterministic relation with performance. The color gradients provide a better quantifiable evaluation than the subjective evaluation and clinical trials.

## 5 CONCLUSIONS

We proposed the entropy of color gradients image (ECGI) criterion to evaluate the performance of colonoscopy. The color gradient image was introduced as the extracted feature under the consideration of the texture, the edge, and the color. Based on this, its entropy was defined to be the ECGI score. Our method pre-processed the LCI-PairedColon database to extracted ROIs. Next, the color gradient image for every sample was calculated and the effect of light-reflection regions was balanced as well. In the end, the ECGI scores of the complemental gradient images were calculated. Experimental results showed that the ECGI scores of LCI images are significantly higher than those of WL in the LCI-PairedColon database. These results confirmed that the proposed ECGI criterion yields a consistent conclusion with clinic colonoscopy during both polyp detection and classification.

The above indicated that the proposed criterion is reliable and superior. Meanwhile, ECGI criterion is more robust than the subjective methods and simpler than the clinical ones. Furthermore, existing data mainly consists of NBI or magnification enhanced images. To our best knowledgement, the proposed ECGI criterion is novel in that it is so far the only study of computer-aided comparison of polyps using LCI.

Some improvements can be conducted in the future work. Firstly, the ECGI criterion can be generalized to other imaging technologies such as narrow band imaging (NBI). Secondly, all the pixels in the colorectal images are considered independent of each other. We can associate the location information with the color gradients of the pixels to further explore the relationship between the color gradient images and the colorectal images.

## Acknowledgements

This work was supported in part by the National Natural Science Foundation of China (NSFC) No. 61773071, in part by the Beijing Nova Program No. Z171100001117049, in part by the Beijing Natural Science Foundation (BNSF) No. 4162044, in part by BUPT Excellent Ph.D. Students Foundation No. XTCX201804.